\documentclass[conference]{IEEEtran}

\usepackage{tikz}
\usepackage{amsmath}
\usepackage{graphicx}
\graphicspath{ {./images/} }

\usepackage{filecontents}
\usepackage{listings}
\usepackage[linesnumbered,ruled]{algorithm2e}
\usepackage{colortbl}
\usepackage{paralist}
\usepackage{circledsteps}
\pgfkeys{/csteps/fill color=black}
\pgfkeys{/csteps/inner color=white}

\definecolor{dgrn}{rgb}{0.0, 0.5, 0.0}
\definecolor{dred}{rgb}{0.7, 0.1, 0.1}

\begin{document}

\date{}

\title{CYPRESS: Transferring Secrets in the Shadow of Visible Packets}

\author{\IEEEauthorblockN{Sirus Shahini}
	\IEEEauthorblockA{University of Utah\\
		sirus.shahini@utah.edu}
	\and
	\IEEEauthorblockN{Robert Ricci}
	\IEEEauthorblockA{University of Utah\\
		ricci@cs.utah.edu}
	}

\IEEEoverridecommandlockouts
\makeatletter\def\@IEEEpubidpullup{6.5\baselineskip}\makeatother

\maketitle

\begin{abstract}

Network steganography and covert communication channels have been studied
extensively in the past. However, prior works offer minimal practical use for
their proposed techniques and are limited to specific use cases and network
protocols. In this paper, we show that covert channels in networking have a much
greater potential for practical secret communication than what has been discussed before.
We present a covert channel framework, CYPRESS, that creates a reliable hidden communication channel by 
mounting packets from secret network entities
on regular packets that flow through the network, effectively transmitting
a separate network traffic without generating new packets for it. CYPRESS establishes
a consolidated decentralized framework in which different covert channels for
various protocols are defined with their custom handler code that are plugged into the system
and updated on-demand to evade detection.
CYPRESS then chooses at run-time how and in what
order the covert channels should be used for fragmentation and hidden transmission 
of data. We can reach up to 1.6MB/s of secret bandwidth in a network of ten users
connected to the Internet.
We demonstrate the robustness and reliability of our approach in secret
communication through various security-sensitive real-world experiments. Our
evaluations show that network protocols
provide a notable opportunity for unconventional storage and hidden transmission of
data to bypass different types of security measures and to hide the source of various cyber attacks.

\end{abstract}

\section{Introduction}
Hiding the communication between two or more parties can be performed for different
reasons, including protecting users from a monitoring adversary or, conversely, hiding 
malicious activity from other users and security controls. Hiding data
can be done through encryption to make the plain data unrecognizable or through
sending data within communication channels that are typically not monitored or
not perceived as regular data carriers. Covert communication in networking has been a
well-known research area using both timing and storage channels
~\cite{cabuk,gianvecchio,cabuk2,lee,chan,chan2,shah,wang,abad2001ip,Dakhane,Giffin,Jones,Luo,rowland}.

A timing channel works by modifying transmission delays
between packets ~\cite{gianvecchio,lee,Shahini} which is measured by the other party
and translated into data bits. On the other hand, storage-based covert channels
use specific fields or gaps either in packets headers or data to
transmit additional data. A simple example of a storage unit is the TCP
acknowledgment number ~\cite{lee}.

While covert channels transmit data using
\textit{legitimate} network packets, these packets must exist in the network in
the first place. This means that in a real-world usage scenario, either
the network packet flow of the existing users should be intercepted and modified, or
the sender must generate their own legitimate-looking packets to use as the
covert channel carrier in which case, the added \textit{unwanted} network traffic can be
enough of a signal to detect an anomaly in the network and expose the covert
channel. This is a significant limitation of much prior
work. In this paper, we particularly focus on transmitting covert data using the \textit{existing}
network packets, not generating new packets to hold the covert data.

Since covert channels usually have an adversarial nature, different mitigation
techniques have been proposed for them
~\cite{xing,Belozubova,chan,Dakhane,Gianvecchio2,Malan,peng}.  To mitigate
timing channels, solutions like adding random noise to inter-packet delays
(IPDs) and statistical comparison of IPDs with a known legitimate
model have been proposed. On the other hand, real-time scanning of packets to
find known patterns or network anomalies can be used for mitigating storage
channels. Usually, each mitigation technique only focuses on one specific
type of covert channel and can be potentially defeated by either using another
form of covert channel or by changing the transmission fields or patterns of
sending covert data. All these techniques are also generally prone to
false positives, which can impede the normal packet flow of a benign connection
by dropping some packets or by necessitating retransmission of packets due to
invalidating some important fields in those packets.

In prior works, the main topic of discussion has generally focused on how to establish or
defend against covert channels. Significanlty less attention has been given to evaluating the
practicality of the proposed techniques. 
Covert channels, by their nature, are limited communication mechanisms. For instance, one
of the primary issues with any covert channel, particularly timing channels, is slow transmission speed. 
The inherent speed limit in the majority of covert channels makes them appear to
have limited utility.

Covert channels are \textit{non-standard} and \textit{unconventional} forms of communication.
Different covert channels use different and non-standard techniques. An important open question in past studies, is how
covert data in the form of normal TCP/IP format, can be scattered over multiple channels
in a way that allows the receiver to correctly recover the original network data. 
 Furthermore, there are no limitations on devising a new covert channel. The
variety and differences among covert channels make it quite difficult to define a
standard strategy for using different (old or new) types of covert channels in one system. 
We will address these challenges for the first time in our study.

From an adversarial perspective, covert communication can be used to maintain
persistence by evading monitoring systems after a successful intrusion. The persistence of
the attackers in the system can greatly increase the collective damage that is inflicted
by the attack, giving the attackers an opportunity to stay in and steal whatever
they need over time. One famous example is the SolarWinds supply chain attack which started
in 2019 and remained undetected for a long time, during which the malicious actors had access to 
the networks of top US security firms. There have also been several successful cyber attacks
against industrial control systems, power grids, water plants, and energy sectors~\cite{Trien}
in the past years.
Software that control industrial infrastructures like SCADA systems, industrial PLCs, sensors, and actuators, are prone to a myriad of vulnerabilities
similar to regular desktop software, and they are even usually easier to exploit after a vulnerability is found~\cite{Gonzalez}.
In many such attacks, it is not the initial breaking into the systems that turns into
a disaster, but the \textit{persistence} of the malicious actors in the network. Maintaining
persistence requires stealthiness and covert means of communication provide an excellent opportunity
to achieve that.
Using covert channels in hacked network equipment is an important topic but it is overlooked
in the literature. In recent years there has been a surge in attacks against
network equipment like home routers and access points
~\cite{wifidef,routermalware,wifihack,routerhack,cvepath}. As an example, in
2017, security researchers reported more than 4.8 million routers were vulnerable
to UPnP NAT injections, and at the time of the report, more than 50,000 devices were
already hacked and were being used to create proxy chains to hide the origin of thousands
of cyber attacks ~\cite{upnp,secretcookie}. A few years earlier, 6 other remotely
exploitable bugs in UPnP implementation had already left more than 80 million routers
exposed to hackers on the Internet ~\cite{upnprapid1,upnprapid2}\footnote{Based on recent
data, as of 2022, some routers with RealTek chipsets are still exploited in the wild
using the UPnP vulnerability.~\cite{Hiesgen}}. In 2019, several serious vulnerabilities were
detected in Broadcom-based cable modems that are widely deployed across the world~\cite{cable-haunt}. The reported bugs
presented an opportunity for a wide range of attacks including remote code execution.

IoT devices, which
are increasing in number with unprecedented speed, have also been one of the
main targets of cyber attacks in the past few years
for purposes like creating botnets, spreading
malware or spying on users~\cite{iotlink,Deogirikar,Haseeb}. These devices can also
turn into covert channels to secretly transfer data between members of a network.
We will demonstrate in our experiments that this covert communication can pose a serious 
challenge for intrusion detection and network monitoring tools~\cite{Bekerman,Prasse,malw1,ahmed,Piskozub} 
to identify malware activity and detect infected systems.

We present CYPRESS, a covert channel framework that sits at a network node that
manages or forwards traffic. This can be a router, a desktop or server machine,
a network device like a WiFi
access-point, cable modem, managed switch or an end device such as a smart TV, camera
or IoT device.  CYPRESS
blends and extracts arbitrary covert network traffic with legitimate traffic
passing through the node, effectively providing a practical hidden
communication channel between multiple devices in a local network or in case of
chaining the gateways, at the Internet level. CYPRESS instances work in pairs. 
We do not focus on any specific form of
covert channel in CYPRESS; covert channels can be defined for the framework and
CYPRESS chooses how to use them based on the network traffic and the characteristics of the
defined channel. We will show how it is
possible to provide hidden Internet access for a secret host using CYPRESS such
that the secret user can use the provided Internet connection like a
legitimate connected device within the network just by piggybacking packets from
other users. Also, we demonstrate how different security measures at the network
level can be bypassed using this technique.

In short, we make the following contributions:
\begin{compactitem}
\item We present a low-level covert channel framework, CYPRESS, that is able to
use and adapt to different types of covert channels and transmit data using
existing network packets (i.e. without generating carrier packets) with an
acceptable speed even with a limited bandwidth. CYPRESS uses a novel strategy for 
scoring and distribution of covert packets on normal network traffic. We devise a
mechanism to split data from a hidden stream across multiple covert channels to
increase bitrate and robustness, and create a synchronization method that allows
the transmitter and receiver to reliably store and extract the hidden stream from
different selected channels.

\item We show that network-based covert channels have a significantly greater
potential to perform hidden communication than what has been demonstrated in the past.

\item We use CYPRESS in multiple real-world experiments to show its potential
and practical use, including:
	\begin{compactenum}
		\item Creating a working secret Internet connection for a secret
		user in a local network.

		\item Bypassing NAT rules to initiate a connection into a
		NAT protected network.

		\item Bypassing IP-based firewall rules to allow connections in
		and out of a protected network for disallowed devices.

		\item Cross access to protected network segments from disallowed
		segments.

		\item Leaking security-sensitive data out of a host through
		legitimate connections made by the host.

		\item Evaluating the resistance of the framework to mitigation
		techniques.

	\end{compactenum}

\end{compactitem}

\noindent \textbf{Ethical Considerations} \\
\noindent Various experiments in this paper discuss stealth data transmission by exploiting
the network communications of other users. The experiments were carried out in a controlled 
environment, and all the test devices were owned and set up by the authors. No human subject without 
explicit participation consent was involved in the experiments. There is no information in the reported 
results that can be used to identify any user or user device.

\section{Overview}
In networking terminology, a gateway is a network entity (hardware or software)
that operates as a \textit{gate} to connect multiple networks.
A network includes one or more hosts. Gateways are
normally able to \textit{route} packets between the hosts
(clients) and forward a packet to the correct next node. Network packets have a layered
structure that is defined in TCP/IP protocol model~\cite{forouzan,fall}
\footnote{We use the 5-layer TCP/IP model for presenting this
work, not the 7-layer OSI model.}.
Each layer has various types of information
about the service that is used in that layer and how to route or manage the packet or how to
parse the data from the higher layers. Network switches work at layer three
and potentially at higher levels. In this paper, we do not
consider an unmanaged layer-2 network switch a gateway.

Whenever a host needs its packet to be routed, a gateway must be connected to
the host. The gateway can be one of the hosts themselves, or it can be a router
device like a home WiFi access point. It is safe to assume that whenever a user's
network packets pass through more than two hosts, there is a gateway to route
the packets somewhere in the transmission path. This includes access to
the Internet and almost all the cases when a device communicates with multiple hosts
within a local private network.

A gateway is an ideal place for centralized monitoring and management of packets
in any network environment. Many commercial network routers have various management features
to enable system administrators to analyze and control the network activities of
the connected hosts. Routers are increasingly targeted by cyber criminals, and
as mentioned before, there has been a significant number of successful
attacks to obtain unauthorized access to router devices which technically work as
gateways in home and commercial networks. A gateway often hides the hosts
behind a NAT firewall and acts as an effective defense against certain types of
remote attacks. This defense, however, can thoroughly fall apart if the gateway is
compromised. 

CYPRESS runs at gateway level, and we investigate the potential of covert
communication between gateways in:

\begin{compactitem}
\item \textbf{providing pseudo-anonymous Internet access to select users inside the
network}
\item \textbf{bypassing security policies and abusing network configurations to
transmit data between hosts in an unconventional manner}
\end{compactitem}

\subsection{A Secret Internet}
Anonymity networks have been one of the main research topics of online privacy.
Different solutions have been proposed and implemented to provide anonymous
network access~\cite{roger,i2p,Chothia}. The Tor
Project~\cite{roger} for example, is the most popular anonymity network with
more than 2 million users per day~\cite{tormetrics}. Encryption is the heart of
anonymity networks; by carrying encrypted user packets through different
intermediary routers, a monitoring adversary cannot directly correlate the
packets to the originating users. However, the user can not hide
the fact that they are using the anonymity network or a proxy server; there is
still a connection to a specific address that can be monitored or blocked.
There are various attacks against anonymity networks to
deanonymize users~\cite{Karunanayake,lin,nepal,jansen} or to destabilize the
target anonymity network to disconnect or interrupt connected
clients~\cite{nasr,Chunxiang,Vaidya,danner}.

We propose a different solution for providing not only anonymous but also secret
connection to the users within the target network in which CYPRESS runs on the gateways: we can
transmit a user's data between gateways by mounting the entire TCP-IP protocol
suite on top of packets sent from other users rather than initiating a new
connection to create an anonymous session. Using covert channels at
different protocols, we mix network packets of select users (\textit{secret
entities}) with regular users (\textit{visible entities}) to create a two-way
normal Internet connection for \textit{secret entities}. From the perspective of
an adversary that is monitoring the packets passing through the gateways, no new
connection of any form is detected to signal the potential use of an anonymous
communication between users or with outside of the network.

\begin{figure*}
\centering
\includegraphics[width=1.0\textwidth]{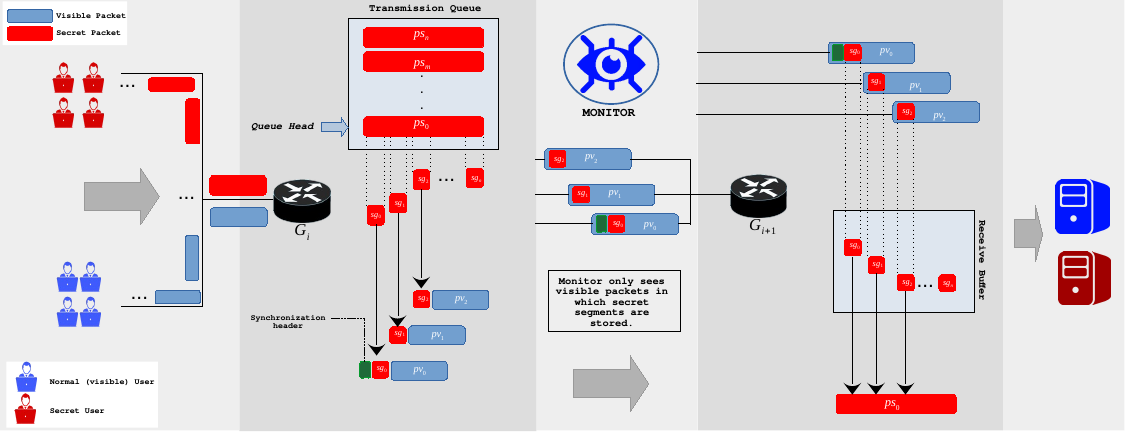}
\caption{High-level view of CYPRESS operation model. Secret packets stored in the transmission queue, are fragmented into secret segments and
transferred to the next gateway. The segments are temporarily held in a local buffer until the original packet can be reconstructed at
the receiver side.}
\label{fig:base_model}
\end{figure*}

\subsection{Evading Security Controls}
In any cyber attack, initial break-in and \textit{maintaining} access are
equally important. Similarly, for the defenders, detecting and stopping a
successful attack after it happens is equally or even more important than
preventing the attack. In the majority of intrusions, the inflicted cost and
damages will be negligible if the attack is detected and stopped soon enough.
Conversely, it can be disastrous if hackers stay unnoticed in the system for a
long time ~\cite{thn,mlwrbytes}. In this paper, we examine the use of covert channels
as an effective lateral movement strategy in a compromised network to hide the intrusion
and exfiltrate data out of the protected network without triggering network
monitoring or intrusion detection systems.

\section{Design}
\newcommand{\chigh}{\textcolor{red}{High}}
\newcommand{\cmoderate}{\textcolor{orange}{Moderate}}
\newcommand{\clow}{\textcolor[HTML]{009f00}{Low}}
\newcommand{\scirc}[1]{{\small\Circled{#1}}}

First, we define the following terms:

\noindent \textit{node}: Any member of our experimental network is a node.\\
\noindent \textit{secret node}: A node that needs to hide its network communication with at
least one other node through CYPRESS.\\
\noindent \textit{gateway}: A node that forwards or routes packets between at least two other
nodes.\\
\noindent \textit{cGateway}: A gateway on which a CYPRESS instance has been installed and is running.\\
\noindent \textit{visible node}: A node that is neither a secret node nor a gateway.\\
\noindent \textit{visible packet}: A packet generated by a visible node used
for normal communication with other nodes or with the Internet. The packet is visible
to monitoring software.\\
\noindent \textit{secret packet}: A packet generated by a secret node.\\
\noindent \textit{monitor}: A node between at least two gateways that can monitor,
drop and manipulate packets.\\
\noindent \textit{protected path}: A physical wired or wireless route between at
least two gateways and \emph{with} a monitor anywhere on the path.\\
\noindent \textit{unprotected path}: A physical wired or wireless route between at
least two gateways \emph{without} a monitor anywhere on the path.\\

Our goal is to hide the communication between secret nodes or between a secret node and
a visible node within a protected path. We assume that CYPRESS runs on the
gateways that are used for our operational stages. CYPRESS has two main operational stages:

\textbf{Fusion}: Packets from secret nodes are fragmented into smaller
segments to be merged into visible packets. Visible packets are then transmitted
to the next gateway.

\textbf{Extraction}: Fragmented segments of the secret packets are extracted
from the received carrier packets, formed into their original structure, and then
transmitted to the next node, out of the current protected path. A carrier packet is
analogous to the carrier wave in radio communication, except that in our model (using the
same analogy), the carrier wave is not generated for modulation, but existing waves 
in the environment are modulated.


In Figure~\ref{fig:base_model}, we have shown a high-level view of CYPRESS
communication model.
We describe each stage separately, and then we present our experiments.

In our threat model, we assume that CYPRESS is installed on the cGateways, either deliberately by the 
network administrator or by an attacker after a successful intrusion. Even if CYPRESS is installed 
on a local machine other than the default gateway, it is possible---depending on the specific situation---
to force other clients to use that machine as their default gateway by techniques like ARP cache poisoning or ICMP redirect
messages to advertise a rogue gateway \footnote{It has been recently shown that even
off-path attackers can manipulate the cache entries of the routing table of a remote target using
ICMP redirect messages~\cite{icmpredir}.}. That said, the goal of this work is to
assess the capabilities of CYPRESS, not its deployment. Further, CYPRESS does \textbf{not} have any
control over the monitors which enforce access controls and network security policies.

\subsection{System Internals and Operation Model}

There are various ways to hide data~\cite{Mileva} in different layers
of network packets that have been extensively covered in prior work. 
While we use different storage candidates in CYPRESS, we will not explain each one separately
as it is not the purpose of this paper (CYPRESS does not use timing channels). 

\begin{figure}
\centering
\includegraphics[width=0.7\columnwidth]{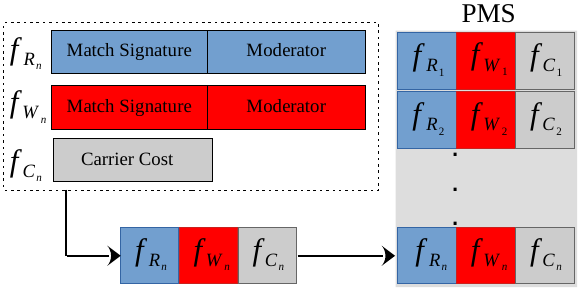}
\caption{Read-Write handler pairs are defined, implemented, and then plugged into the PMS for each
new carrier candidate.}
\label{fig:hnd_pluggs}
\end{figure}

We define
each potential storage covert channel that we want to use in covert
transmission, using low-level packet handler functions. Each handler function
comprises a \textit{match-signature}, a \textit{packet-moderator} and a \textit{carrier-cost}. The match-signature is 
used to describe the packets that the handler function supports. Packet-moderator is the actual code that manages how the packet should be
modified or read to store and extract data respectively 
and the \textit{carrier-cost} is a number that is assigned
to the handler which shows how the cost of this carrier is compared to other carriers. The cost
is determined based on the synchronization overhead and the experimental results of transmitting 
data using the carrier (refer to Appendix~\ref{ap:hnd_example} and Appendix~\ref{ap:costs} to see examples
for handler functions and cost assignment). Handler functions are written in writer-reader pairs. 
For each writer, there is exactly one reader function that knows how to extract data, defined along with its
corresponding writer function in the sender cGateway.  
Each handler function is pluggable into the main
framework code (i.e. the framework core). In Figure~\ref{fig:hnd_pluggs} we have shown a simple model of adding new functions to the
CYPRESS core. The set of handler functions is called the Packet Moderation Set (PMS). 
Each time that CYPRESS
needs to check whether a packet can be used as a carrier, it consults the PMS. If more than one potential 
handler is found, one is used based on the defined cost ($f_C$) and the current state of the network flow. We generally
avoid using more than one storage candidate for the same packet.
In Table~\ref{tab:candidates} we have listed some examples of candidate storage fields. 
While there are tens of other potential candidates in layer 3 and 4 from different protocols (like IPv6, DNS and
IGMP),
we chose a select set of candidates to execute our tests for two reasons. First,
evaluating individual storage channels and comparing them is not the purpose of this
work. Second, increasing 
the number of these candidates does not necessarily improve the covert communication.
We have designed CYPRESS to maintain network stability and speed. Visible users should
not suspect their network communication is being tampered with or interrupted. Increasing the
number of possible candidates makes attaining this required stability significantly more
difficult. When data is distributed on the carrier packets, it is clustered in multiple segments
based on the type of the carrier channel. Each segment will belong to its own carrier type. When
CYPRESS switches to another carrier channel (e.g. from TCP sequence number to ICMP payload)
it adds \textit{synchronization headers} to the beginning of the new segment and writes additional
data along with the secret data to recover destructive packet fields. A storage field is destructive if it has
to be recovered to its original state before leaving the last cGateway and delivered to a regular
gateway to prevent the destination from dropping the packet. A checksum value is an example of a destructive field.
We will present more details in the next sections.

\begin{table}[t]
\small
\centering

\caption{Some sample storage candidates}
\label{tab:candidates}

\begin{tabular}{|c|c|c|c|}
\hline
\rowcolor{gray!30}
Protocol & Candidate & Size (bits) & Cost \\
\hline 
\hline
IPv4 & IP identification field & 16 & \chigh\\

\rowcolor{gray!30}
IPv4 & IP fragmentation offset & 13 & \chigh\\

IPv4 & IP options & [0,320] & \cmoderate \\

\rowcolor{gray!30}
IPv4 & Individual options & Varies & \cmoderate \\

IPv4 & Header checksum & 16 & \clow \\

\rowcolor{gray!30}
IPv4 & Time to live field & 8 & \cmoderate \\

IPv4 & Flags & 3 & Varies \\

\rowcolor{gray!30}
TCP & Sequence number & 32 & \chigh \\

TCP & Acknowledgment number & 32 & \chigh \\

\rowcolor{gray!30}
TCP & Window size & 16 & \cmoderate \\

TCP & Header checksum & 16 & \clow \\

\rowcolor{gray!30}
TCP & Options & [0,320] & \cmoderate \\

ICMP & Data payload & Varies & \clow \\

\hline

\end{tabular}

\end{table}

\begin{figure}
\centering
\includegraphics[width=0.8\columnwidth]{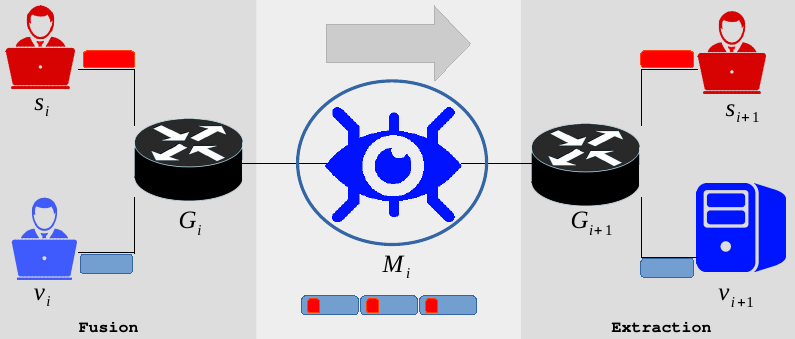}
\caption{High-level view of Fusion-Extraction flow.}
\label{fig:base_simple}
\end{figure}

When a packet is received at the cGateway, CYPRESS checks the network and
physical addresses of the packet to specify whether the packet should be used for fusion (i.e.
as a carrier packet) or for extraction (i.e. it has secret segments).  

\begin{figure}[h]
\centering
\includegraphics[width=0.6\columnwidth]{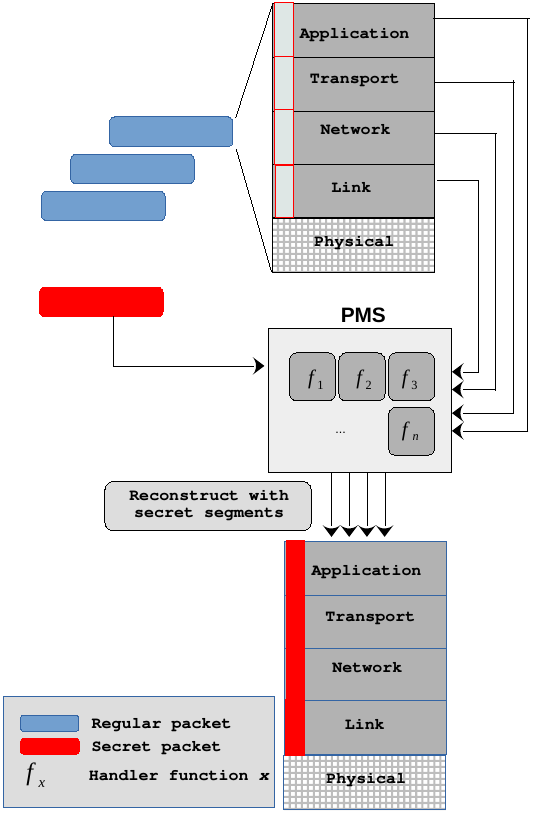}
\caption{For each potential carrier packet, at least one handler function is chosen for handling 
secret segments.}
\label{fig:handlers}
\end{figure}

Figure~\ref{fig:handlers} shows the operation model of CYPRESS for data storage (fusion).
The carrier gaps in the network packets are filled using the handler functions in the PMS. The goal is 
to embed secret traffic in the packets while at the same time maintaining the correct structure of and the 
original data carried by the input packets.

\subsection{Fusion}

When a packet is chosen for the fusion stage, a storage function is selected
to handle the segmentation and storing the segments based on the capacity of the
covert channel (for example, the size of the fields in which the segment is being
stored) and writes the segment at the target location potentially with a
synchronization header. A synchronization header is used to synchronize gateways
and restore the segments to reconstruct the original secret
packets in the extraction stage.
In Figure~\ref{fig:base_simple}, \textit{Fusion-Extraction} has been shown in
a simple network configuration.  In
the fusion stage, packets from the secret and visible nodes arrive at cGateway
$G_i$. For the sake of simplicity, here we assume only one secret node $s_i$ and
one visible node $v_i$ are directly connected to $G_i$.  Further, $G_i$ is
connected to the cGateway $G_{i+1}$ through monitor $M_i$. We want to hide the
network traffic of $s_i$ from $M_i$ and smuggle packets generated by $s_i$
past the monitor.  While we describe only one direction of
covert transmission, CYPRESS has a symmetric design, and the following procedure
can be used for both directions.

When $G_i$ receives a packet from $s_i$, it stores the packet in a local
transmission queue to be later
sent to $G_{i+1}$ through $v_i$ packets in small segments whose
characteristics depend on the visible carrier packet. When $G_i$ receives a
packet from $v_i$, it checks the packet structure, makes the next
segment accordingly from the transmission queue, stores the segment on the candidate storage
locations and routes the packet to $G_{i+1}$. CYPRESS can encrypt the segment
before transmission, but as we show in Section~\ref{sect:detection_test},
even without encryption, various security solutions fail to detect malicious traffic
that is transmitted as secret segments between gateways.

Carrier packets are not necessarily consecutive. When a packet should not be used as a carrier, CYPRESS \textit{excludes}
the packet by inserting a specific synchronization header (an exclusion signature), instructing the receiver that this packet does not contain
secret segments or internal management information. The exclusion might happen because the packet 
is not useful for carrying secret segments for having a bad size or because the collective
cost of synchronizing the current segment exceeds the acceptable limits. \textit{Exclusion signatures} do
not add network overhead in the transmission. 

\subsection{Extraction}

The extraction stage is similar in choosing carrier packets
and handler functions, but performs a reverse operation to load the secret segments and rebuild them into the
original form.
In the extraction stage at $G_{i+1}$ the reverse operation is performed. When a
packet that is originating from $G_i$ arrives, the packet structure is investigated, and
if it contains a secret segment, the segment is extracted, decrypted (in case of
encrypted segments) and stored in a local buffer which
will hold all the segments of a secret packet before the packet can be
completely recovered. The visible packet originating from $v_i$ is normally
routed to its destination outside $G_{i+1}$. When all the necessary segments to
form an original secret packet are received at $G_{i+1}$, it reconstructs the
secret packet and routes it to its correct destination, which in this case is
$s_{i+1}$. While the monitor $M_i$ only sees packets originating from $v_i$, we
effectively carry secret packets of $s_i$ through the same physical channel that
visible packets from $v_i$ are sent through.

In Appendix~\ref{ap:alg} we have provided the core
algorithms of the fusion and extraction stages in CYPRESS.

\subsection{Operation Assumptions and Limitations}

The cGateways that take part in transmitting secret segments do not have to be 
adjacent nodes. CYPRESS does not manipulate routing information that is necessary
to forward packets and a modified packet will not be rejected by a regular gateway.
Routers are not interested in evaluating the packet fields that are used by CYPRESS
for covert data transmission.

To maximize the 
transmission capacity, CYPRESS does not perform an integrity check for the assembled secret segments. 
That is, we do not insert extra bits specifically for integrity checks. It will be up to the receiver side to check the integrity of the secret
packets \textit{after} extraction is completed. This will normally happen on the receiver side at the OS or application level. Corrupted
packets will result in retransmissions through CYPRESS.
Further, we assume that all carrier packets are sent and received in the same order. Two cGateways that directly
interact with each other are normally in the same network segment. Although the cGateways do not need to be adjacent,
the two communicating cGateways are close to each other. In our tests, there are not
more than two nodes between a cGateway pair that are engaged in a fusion-extraction session. That said, 
there can be multiple pairs within a protected path. Packet loss and reordering did not happen in our
experimentes. But both packet loss and reordering can be handled by extending the synchronization header functionality
and adding more sync codes to the system.

\subsection{Packet Manipulation Challenges}

When we overwrite any part of a carrier packet to add secret segments, one of these three cases might happen:
\scirc{1} It does not invalidate the packet and does not affect the quality of the
	service on top of which we are mounting our secret data. The packet still conforms to the structural 
	protocol definition of the original packet and will not be rejected by the destination machine
	or non-cGateways along the route. TCP initial sequence number is an example of this type of
	carrier fields.
\scirc{2} It does not invalidate the packet but might affect the quality of service and
	the stream of packets in general. The \textit{options field} in layer 3 or layer 4 is an example of this type of carrier
	fields. Overwriting options does not necessarily corrupt the packet but might affect the 
	connection speed and the stability of the network communication.
\scirc{3} It invalidates the packet and the packet cannot be processed regularly in
	a non-cGateway or the destination machine. Overwriting any header checksum value 
	(in the network or transport layer) or modifying
	the destination addresses are examples of this type of carriers. We call these 
	storage candidates destructive candidates. Note that, as we mentioned previously, CYPRESS
	does not use destructive candidates that manipulate routing information that is required
	by regular gateways.

Defining candidates requires manual effort. There is no general rule to determine or predict how modifying 
different packets' fields affects the overall network stability. This is completely dependent on the specific protocol and the type of the 
carrier packet. To add a new handler, we should carefully study the structure of the corresponding packet and the roles
of its various network fields whose details are not relevant to this paper and 
the design of CYPRESS.

\subsection{Packet Recovery Challenges}
For any carrier field that we overwrite, there are three cases regarding the recovery of the original 
value of the field: \scirc{1} The original value does not need to be recovered. For example,
reserved fields, debugging or network routing information can in some cases, be safely overwritten
without corrupting the packet or having the destination reject the packet. \scirc{2}
The original value can be recovered or corrected by any node in the route without needing 
additional information about the packet or the modified field. For example, an overwritten checksum value
can be trivially corrected without knowing what the original value was. \scirc{3}
The original value can only be recovered or corrected if additional information is communicated
between the two parties. An example is the sequence numbers of TCP packets. If the initial sequence number is
overwritten, any node that wants to recover the original value needs the value to be communicated from an external node
that has recorded the initial value. We refer to this issue as \textit{augmented correction}. Augmented correction
happens when it is both \textit{necessary} to correct an overwritten network field in a carrier packet, and we \textit{must} have additional information
about the packet from at least
one external cGateway to correct the modified field. 

Generally, the type and characteristics of a carrier candidate specify how complex it is to use it as a covert channel and
how it can negatively affect the overall covert communication stability. Candidates that need augmented correction along the
way drastically slow down the covert communication. An exclusion signature is not written on any part of a packet that 
might create augmented correction
and it will not cause the receiver side to update the internal structures that define the state of the current secret segment
being extracted.

Whenever we want to choose a network carrier we have to ask two questions: Does it introduce any changes to
the carrier packet that needs correction in at least one of the next cGateways before being delivered
to the final destination? And if it does, what are the costs of recovering/correcting the affected
network fields? If it does not, to what extent the hosts that receive, process or route the packet can tolerate
the changes that are applied to the packet network fields by the cGateway? We have observed that this tolerance
level is quite different for different carrier candidates.

Table \ref{tab:candidates}, shows the relative cost of using each of the tested
candidates. The cost comprises the synchronization overhead, the necessary additional information
to send along with the secret data and the potential of interrupting the stream of packets. These factors
translate into the total time that is spent on completing a secret communication session and the
slow-down that happens for visible users' connections. Additional
information is what CYPRESS needs to recover destructive fields to their original value or to a structurally
correct value to prevent any non-CYPRESS node from dropping the packet. Network interruption
occurs for various reasons. The more interruption we have, the less stable the connection will be
for the visible users. Creating a balance between the speed and reducing interruption and the overhead of managing 
different types of candidate carriers is a major challenge in CYPRESS.

The selection of handler functions is managed based on four metrics: \scirc{1} \textbf{Cost}: CYPRESS 
prioritizes handlers that impose a smaller cost on the system. To reach a stable
secret communication we minimize cost in terms of synchronization complexity and sustaining the correct
structure of the carrier packets based on the respective protocols that define the packets; hopping back and force on different carrier 
channels requires synchronization, and overwriting 
certain parts of a packet has a higher chance of structurally invalidating the packet. Writing on IPv4 options field for example,
is prioritized over writing on IPv4 identification field.
\scirc{2} \textbf{Recovery requirements}: Recovery means modifying the carrier packets after the \textit{extraction} stage
in another cGateway to overwrite the affected packet fields due to secret data storage. As described before, recovery is not always necessary and when
it is necessary, its complexity varies on a case-by-case basis. CYPRESS prioritizes handlers that do not require augmented 
correction. Augmented correction requires sending additional recovery data along with the secret chunks, significantly
slowing down the secret communication.
\scirc{3} \textbf{Carrier size}: Carrier candidates that provide bigger capacities to transfer secret chunks are preferred.
For example, IPv4 header checksum is preferred over IPv4 time-to-live field.
Since CYPRESS only uses the \textit{existing packets} from visible users to transmit data, it tries to transmit 
as much data as possible using a packet to reduce the total number of affected packets.
\scirc{4} \textbf{Number of fusion operations}: Increasing the number of fusion operations in one packet, drastically 
complicates the synchronization. In other words, CYPRESS prefers to use one handler for any chosen carrier
packet. This means that when different candidates can be used, we preferably limit
the number of chosen handlers for a carrier packet to one.


\subsection{Encryption and key negotiation}
If encryption is enabled between two communicating cGateways, this is how the gateways
create and share their keys: each cGatewasy generates a 2048 bit RSA key pair. Each cGateway
transmits to the other cGateway, its public key and the MAC address of the NIC it is reading packets
from. The numeric representation of the right-most 32 bits of the MAC address is used to specify
which of the two gateways should generate the symmetric key. A gateway with a greater value of this
number will generate a random 16-byte symmetric key, encrypt it with its peer's public key and send it to 
the second gateway. After this, all secret segments -and not the synchronization headers- 
will be encrypted using AES256-CBC. Synchronization headers cannot be encrypted.
Furthermore, the key transmission itself happens in secret segments through visible carrier packets.

\subsection{Example}

We use a simple example of embedding a piece of secret data in five packets ($p_1$ to $p_5$) passing
through a cGateway when augmented-correction is not required. $p1$, $p2$, $p4$ and $p5$ are non-SYN TCP 
packets while $p3$ is an ICMP request
packet. The packets are received and transmitted in the order they are numbered.
Starting with $p1$, CYPRESS consults the PMS to choose a handler for this packet. For this example,
we assume that the handler $f_t$ is chosen to modify the packet, and the carry-target \footnote{The part of
the packet used by a handler to read/write secret segments.} for this handler
is TCP options field. The Match-signature in $f_t$ checks whether the service type of the IP header equals 6 (TCP protocol)
, and if it does, it then checks whether the SYN flag is NOT set in the TCP flags. Match-signatures in different handlers
check the input packet differently and independently; while in this example $f_t$ is matched against only two 
potential fields (IP service type and TCP flags), other handlers might have more or fewer checks to match.

\begin{figure}
\centering
\includegraphics[width=1\columnwidth]{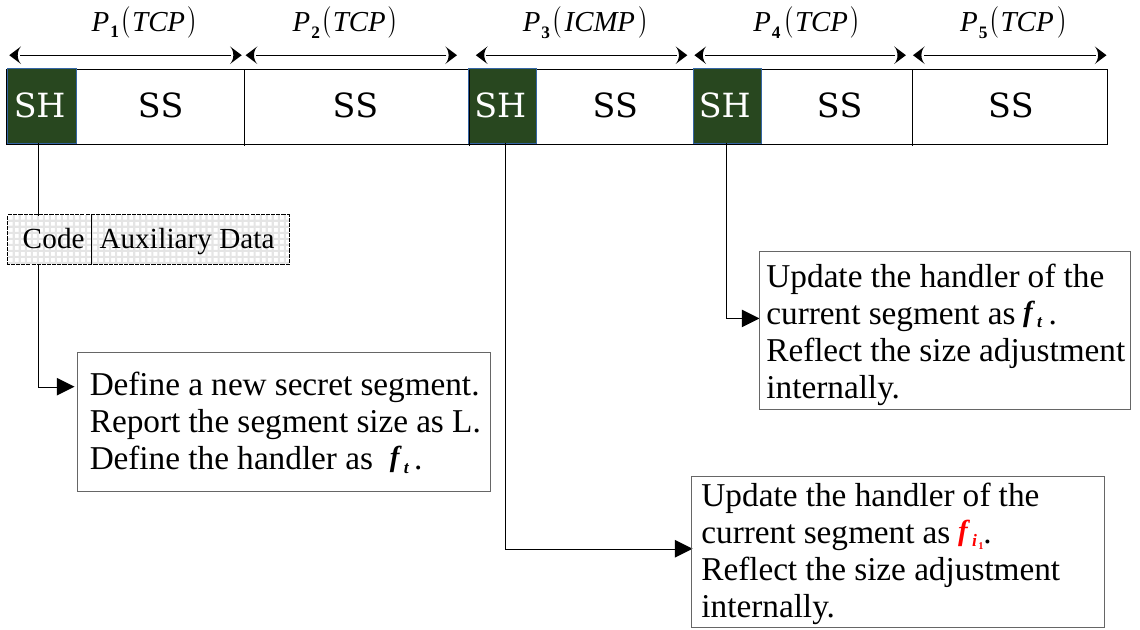}
\caption{The sequence of the extracted secret bytes carried by $p_1$ through $p_5$. 
SS and SH are secret segment and synchronization headers respectively. The synchronization headers
have been transferred by the first, third and fourth packets.}
\label{fig:example_pcks}
\end{figure}

The TCP protocol allows up to 40 bytes of TCP options to be transmitted in a single
packet. Assuming $p1$ is the first packet that carries secret data, a synchronization header must be prepended
to the secret segment prepared for this packet. We set the default size of the synchronization header to 3 bytes including 8 bits of synchronization
code ($2^8=256$ total synchronization codes) and 16 bits of synchronization data. The synchronization header
instructs the next cGateway where and how to extract the secret segment. This will leave us with up to 37 bytes of
secret data to be transmitted using $p1$; a secret segment of 37 bytes is packed into the TCP options of $p1$. Then
$p1$ is routed to the next gateway. Since $p2$ has the same signature as $p1$ (defined as a non-SYN TCP packet), the same handler ($f_t$) is used to add the 
next secret segment to $p2$, this time without a synchronization header. A secret segment of 40 bytes is packed and 
inserted into $p2$ and $p2$ leaves the gateway. $p3$ is a different type of packet and has a different signature to
scan the PMS; an ICMP request packet. Now two cases can happen:

\begin{compactenum}

\item There is only one handler, $f_i$, for this signature returned from the PMS. A synchronization header
will not be necessary and the secret segment is directly embedded into the packet. Assume that $f_i$ uses ICMP payload
as the carry-target. The default ping program in Linux writes 56 bytes in the payload of ICMP request packets.
This usually includes an 8-byte timestamp and 48 bytes of arbitrary data. A secret segment of 56 bytes is packed
and written into $p3$ which is then rerouted and leaves the gateway. For $p4$ and $p5$ the same procedure as $p2$ will be
executed, and at the end we transmit one 40-byte secret segment in each of them. Collectively, we will have 
transmitted 213 bytes of secret data at the moment $p5$ leaves the current gateway.

\item There is more than one handler for $p3$. We assume that $f_{i_1}$ is chosen to handle 
$p3$. In contrast to the previous case, we have to prepend a new synchronization header to instruct the next
cGateway what handler to use for extraction. If $f_{i_1}$ uses ICMP payload to write data, we pack a 53-byte ($56-3$)
secret segment and route $p3$ out of the current gateway. Since $p4$ has a different signature, again we have to synchronize
the current and the next gateway with another synchronization header, instructing the next gateway to expect the next secret
segment from a TCP (and not an ICMP packet). As a result $p4$ and $p5$, using $f_t$, will carry 37 and 40 bytes
of secret data respectively, adding up to total 207 bytes. In Figure~\ref{fig:example_pcks} we have shown a high-level
demonstration of the extracted segment along with the headers.

\end{compactenum}

The extraction stage will reverse the aforementioned procedure to extract the transmitted secret data. The major
difference is that the extraction phase has to \textit{synchronize} itself with the fusion phase; the extractor
scans all the potential packet fields that might contain a synchronization header. After reading the synchronization headers, 
the extractor figures which packets contain secret segments (packets without exclusion signature) and how it should process them. To make it more difficult 
for an adversary to see and recover the secret segments, the PMS has a dynamic nature. The location and the size of 
secret segments for each handler can be modified non-deterministically each time the PMS is compiled. This information can
also be carried through the synchronization headers at the expense of significantly reducing the speed. In our tests, we define
the sizes and locations of secret segments for each handler statically at compile time. 

The given example also shows
how the complexity of managing the packets increases when we increase the number of carrier candidates.
Note that if CYPRESS wants to recover the affected packet fields in this example, the recovery information should also be transmitted
along with our secret data within the secret segments. But normally, recovering ICMP payload and TCP options is not 
required (although we can configure CYPRESS to recover those fields too on the receiver side).

Each secret packet is transmitted independently from the previous and next secret packets. However, the transmission
of each secret packet is handled in a stateful manner. Both the transmitter and the receiver must remember what segments
have been transferred and how to merge the segments after all segments are available at the receiver side.\\

\section{Experiments}
We implemented an experimental prototype of CYPRESS in C under Linux. We compiled 
CYPRESS for i386, x86-64, ARMv7 and ARMv8 architectures used in
different machines in our test environment. PMS which includes the moderator code and match-signature
of handlers has also been written in C. Technically,
it is possible to use high-level languages to implement the PMS with some performance penalty which can be 
a significant concern if CYPRESS is going to run on routing hardware with limited processing power.

We describe the use of our proposed framework in multiple examples, starting from
the base usage, which is \textit{providing a secret Internet connection for a secret user}.
For all the experiments, the target network comprises various connected devices in a laboratory. 
To ensure the flexibility of CYPRESS in regard to the configuration of the test network, we run our tests
in different network topologies. A network configuration file that defines the neighboring nodes for a cGateway
is given to CYPRESS as input upon starting the gateway code. CYPRESS reads the configuration to figure
out where secret users and other cGateways reside. Next, depending on the 
type of the experiment, CYPRESS starts to hide the packets of the secret users inside the network activity of visible users,
passing them from one cGateway to another. cGateways in our experiments are Linux-based x86 and ARM machines.

\subsection{A secret Internet}

In this experiment, we mount the whole Internet traffic of a secret user on the network packets of
at least one visible user. The network configuration is similar to Figure~\ref{fig:base_simple} but
the secret user $s_{i+1}$ does not exist; packets sent from $s_i$ are not limited to one specific secret user. The protected
path in this experiment is $G_i \leftrightarrow G_{i+1}$.
Since the transmission of secret packets depends on the existence of visible packets, the connection speed of
the secret user is limited by the network activities of the visible users. The more visible users we have in the 
target network, the faster
the connection we can provide for the secret user. The visible users in our tests interact with the Internet, similar
to a real user that uses different Internet services for their daily online tasks. 
Note that a secret user does have a valid local
IP address that must be known to the cGateway that is directly connected to the secret user. This IP address
is assigned either statically or by DHCP packets that are either transmitted through an 
unprotected path or carried through the secret segments in a protected path.

First, we evaluate the speed and usability of the provided secret Internet
based on the number and the activity of visible users. Note that covert channels normally are quite slow; a covert
channel is not supposed to be used for network-intensive tasks like downloading or streaming video. The transmission capacity of
covert channels is generally limited to slow secret communication. That said, in our framework, we can reach an acceptable 
\textit{secret} bandwidth for at least one secret user depending on the number of visible users behind the cGateway, the bandwidth used
by the visible users and their network activity.

\begin{figure}[h]
\centering
\includegraphics[width=1\columnwidth]{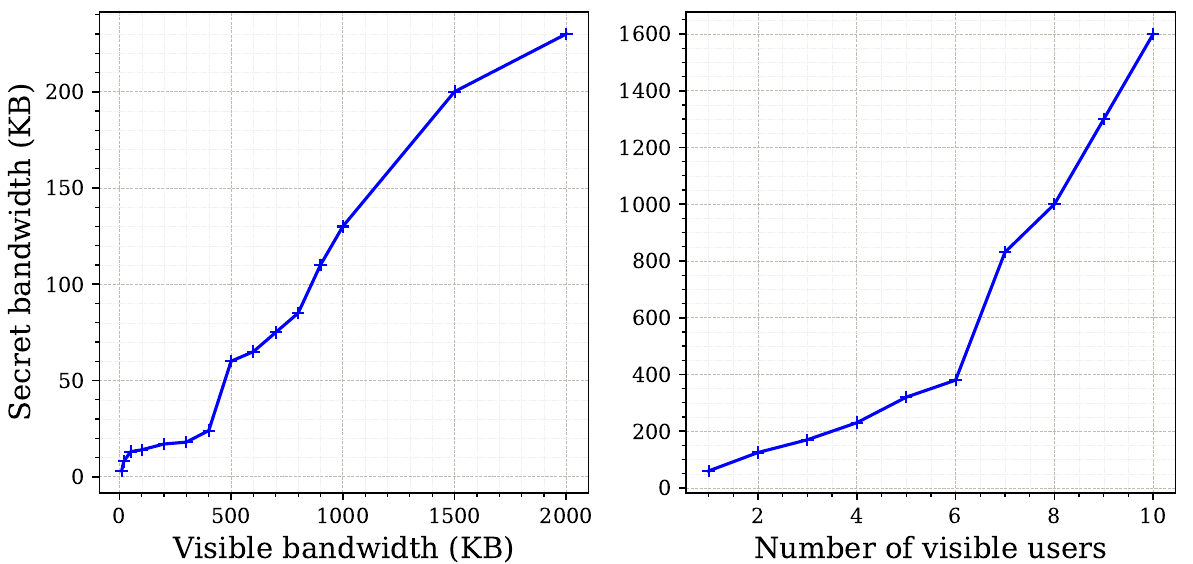}
\vspace{-0.2in}
\caption{The ratio of the bandwidth of the secret Internet to \textbf{a)} the visible bandwidth and 
\textbf{b)} the number of secret users.}
\label{fig:spdbw}
\vspace{-0.15in}
\end{figure}

In Figure~\ref{fig:spdbw} we have summarized the capacity of the provided secret Internet based on the visible bandwidth of one 
secret user (experiment \textbf{a} on the left)
and the number of users (experiment \textbf{b} on the right). In experiment \textbf{a} there was only one visible user
connected to our gateway, and in the experiment \textbf{b} we fixed the bandwidth to 500KB/s for all connected visible users. 
We connected one secret
user to the evaluated cGateway, and we ran a
script on each visible user's machine to consume the whole allowed bandwidth 
by using different popular Internet services.

\begin{figure}[h]
\centering
\includegraphics[width=0.6\columnwidth]{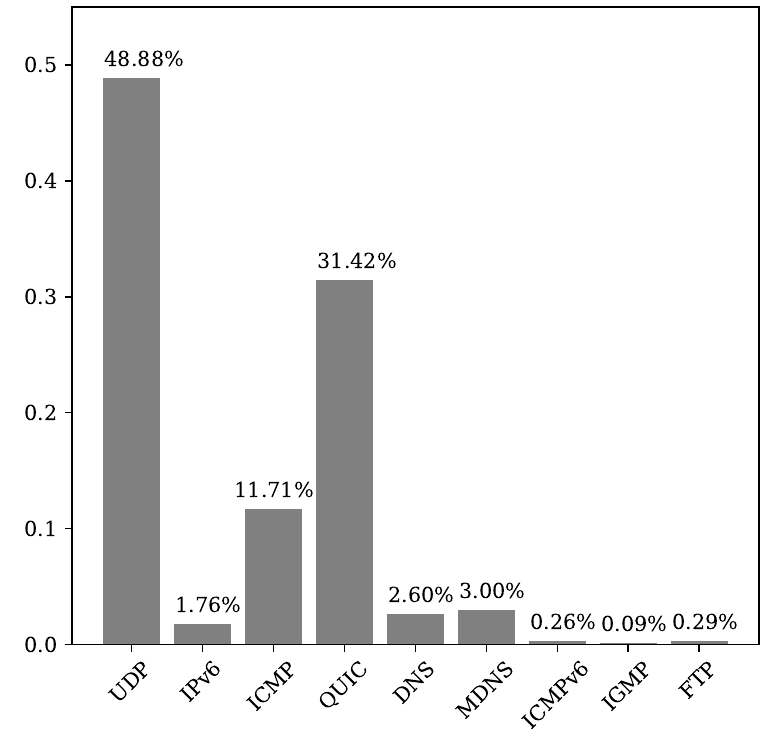}
\caption{Bandwidth usage by different protocols.}
\label{fig:bw}
\end{figure}

The assigned bandwidth to each visible user was used to transmit various types of packets
from different protocols. While there are a handful of different network layer protocols, the absolute majority of
all network packets for an average user in normal situations are IPv4 packets. The majority of these
IPv4 packets have TCP as their transport service in layer four. Apart from these two popular
network and transport layer protocols, there have been other protocols in all
our tests to create a realistic network configuration that happens in normal real-world usage in a user's
machine. We had each visible user perform tasks that are normally done by users when
they use the Internet. In Figure~\ref{fig:bw} we have shown the average percentage of the bandwidth that was used
for different protocols other than IPv4 and TCP. Note that packets might (and usually) have more than one storage candidate
(and hence multiple potential covert channels). For example, an ICMPv6 packet is also a IPv6 packet. In these cases,
the packet is counted in more than one category that is represented in Figure~\ref{fig:bw}. For the TCP case, about $31\%$
of the packets were HTTP, and about $15\%$ were TLS packets.

\begin{figure}[h]
\centering
\includegraphics[width=0.6\columnwidth]{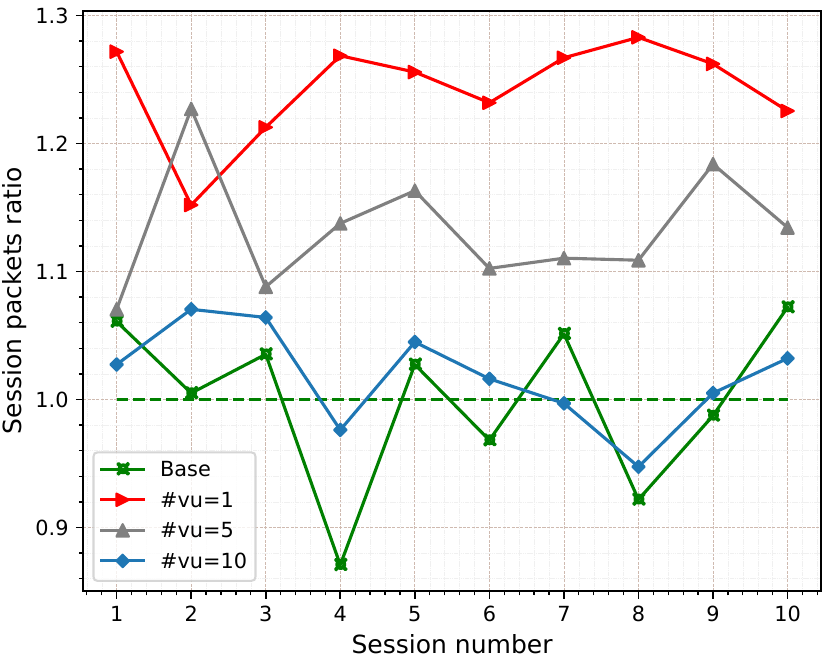}
\caption{The distance between the normalized packet ratio of each covert test case with a constant number of visible users
and the base non-covert transmission session. The base is the ideal case. Lower numbers are better and show
fewer retransmissions with $vu\#=1$ being the worst case. vu\# means the number of visible users in the session. All numbers are normalized
against the average of the packet numbers in the base experiment (green dashed line) pivoted on $y=+1.0$. }
\label{fig:stability}
\end{figure}

As shown in the graphs, increasing the number of visible users improves the 
secret bandwidth more significantly than increasing the bandwidth for each visible user. 
With ten visible users, we could provide more than 1.6MB/s 
bandwidth for the secret user, fast enough for watching a high-definition live stream video without interruption. 
These numbers belong to each 
cGateway, and if we have more secret users behind each cGateway, the 
secret bandwidth will be shared among the secret users. The non-linear speed improvement in these
tests is explained by the fact that network stability is significantly improved when we distribute
the secret payload on a larger number of visible packets. To better show this, we executed another
40 transmission sessions, keeping the bandwidth and the secret payload constant and compared the average number of packets
sent and received at the secret user side (Figure~\ref{fig:stability}).
By increasing the number of visible users,
we gradually achieve a similar network throughput as the base session which is a direct non-covert transmission 
to the gateway used as the best-case scenario; meaning that the plots that are closer to it, show more
stable connections to the cGateways. 
In this test, we only used the comparative representation as the exact packet number in each session
does not matter.

This strategy of carrying a secret payload through other users' traffic can also be used to leak information out of a private segment of the network. 
Information leakage is a serious security risk. While not itself considered a security breach, it can lead
to the full compromise of the system by helping attackers to subvert the defensive measures and form a working
plan of attack. There are different scenarios in which
secret information can be leaked to unauthorized parties, including software bugs or an insider attack. Irrespective of
the source of the leakage, using this covert model, information can be transferred out through a secret connection on top of existing
legitimate connections, as discussed in the experiment.

Note that the reported speed of the secret Internet is achieved by only using the third and fourth network layers of the carrier
packets because we wanted to focus on carrying data through the layers that are \textit{not} designed for
data transmission so that we show how network management fields can be exploited
to carry large amounts of data. Some application layer protocols like HTTP provide a big
opportunity for side channel data transmission and consequently increasing the speed if we use the application layer
for transferring secret segments. But the current implementation of CYPRESS,
does not touch the application layer.

\subsection{Bypassing NAT}
Network Address Translation (NAT) is used in routers
for IP sharing and protecting a local network from specific types of intrusions.
While users within the local network can initiate connections to other nodes in the network
and to the addresses \textit{outside} the local network, unsolicited connections from outside are not allowed. That is,
an external host cannot start a TCP connection into the network. Similarly, UDP packets destined to unseen
ports will be dropped.

\begin{figure}
\centering
\includegraphics[width=0.8\columnwidth]{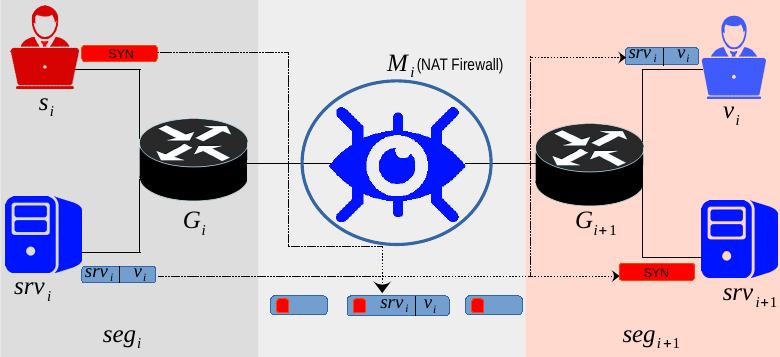}
\caption{Creating a secret connection to $seg_{i+1}$ which is behind the NAT $M_i$.}
\label{fig:nat}
\vspace{-0.15in}
\end{figure}

Using CYPRESS, we demonstrate how we can bypass NAT and successfully establish a connection to a listening
host within the target network. 
We use TCP in this example, though a similar strategy applies to UDP as well. Normally SYN packets from external hosts
are dropped at the NAT. 
In this experiment (Figure~\ref{fig:nat}) we have at least one visible host ($v_i)$ 
in the NAT-protected
network segment $seg_{i+1}$ connected to a cGateway ($G_{i+1}$), and one of the hosts in $seg_{i+1}$ is 
listening for incoming connections (which is not allowed at the network level).
In this example, secret host $s_i$ wants to establish a connection to server 
$srv_{i+1}$ (at the right in Figure~\ref{fig:nat}). $M_i$
performs the NAT operation for $srv_{i+1}$;
incoming TCP connections to $seg_{i+1}$ are blocked by $M_i$.
We call the listening
host the \textit{hidden server} which can be any of the hosts within the protected network. The secret host which
tries to connect to the server is outside the protected network connected to another cGateway ($G_i$) at the other
end of the protected path. Connections to the server $srv_i$ are allowed by $M_i$. This is a common scenario in home networks.

Visible user $v_i$ tries to make a connection to outside; in this case to server
$srv_i$ in segment $seg_i$. We take advantage of this connection and mount the SYN packet from the secret host on top of the \textit{response} packets
from $srv_i$ to $v_i$. The SYN packet is then extracted and sent to $srv_{i+1}$ by $G_{i+1}$, effectively starting a TCP handshake
between $s_i$ and $srv_{i+1}$. From the viewpoint of $M_i$, this new connection to inside $seg_{i+1}$ is not visible; $M_i$ only
sees request and response packets between $v_i$ and $srv_i$. The rest of the secret connection between $s_i$ and $srv_{i+1}$ will
be similarly handled through $G_i$ and $G_{i+1}$, creating a two-way normal TCP connection between the two parties, bypassing the NAT
mechanism implemented in $M_i$.

\subsection{Bypassing IP restrictions}

Traditional ways to monitor packets will not suffice to detect, log or monitor the secret
packets transmitted by our framework. The reason is that we exploit the standard structure of legitimate packets
that are allowed to pass through gateways in the target networks. The covert channels used to transmit 
secret segments do not change the normal structure of the carrier packets while they hide the presence
of new packets from the monitoring software, including Linux netfilter, \texttt{iptables}, 
various IP-based firewalls, and malware detection tools.

\texttt{iptables} is a popular and powerful user-space Linux utility to filter IP packets
based on user-defined policies and rules. Any \texttt{iptables} rule that matches user-defined
values against different fields of the investigated packets (like source IP address, port number and other protocol-specific fields)
will fail to identify and log secret packets transmitted through CYPRESS. 
In the previous experiment, adding
manual iptables rules on the Linux machine used by the monitor to block packets to/from the secret user's IP address
(which is normally not exposed to the monitor)
does not have any effect on the secret transmission. 
Similarly, other firewall solutions
that simply rely on studying those fields to monitor packets will only be able to see carrier packets to/from
visible connections for the same reason. This also implies that we can carry any disallowed payload using the carrier
packets, including a malware executable or exploit code. In Section~\ref{sect:detection_test}, we will present the
result of testing different network monitoring and security software in the next section.
As an extension of this experiment, we have explained in Appendix~\ref{ap:hijack} how we can use CYPRESS to hijack other users' IP addresses
by switching the roles of a secret and a visible user.

\subsection{Cross-Access to Disallowed Network Segments}
Network segmentation or segregation are effective security measures to limit the cost and scope of a successful intrusion
in the network ~\cite{Yuri,segl1,segl2}. In the absence of segmentation, a single intrusion can create widespread access to the entire
network for the adversaries, while with segmentation in place, the attackers are contained within the compromised subnet.
Network segmentation is also used to limit the access of legitimate users to network resources based on custom zone-specific policies
to prevent insider attacks or inadvertent damages to critical parts of the network.
While network segmentation can be implemented at different (including physical) layers, it is most commonly done at layer three.

There are different approaches to implementing network segmentation. For example, managed switches or routers can divide a network 
into multiple segments to enforce custom access policies among the users 
within each segment. Virtual Private Networks (VPNs), virtual machines and host-based firewalls can also be used 
to execute the same task. Whatever technique is used to implement segmentation, its goal is to restrict network traffic 
of any given host to a list of allowed segments. Using CYPRESS we can create a cross-segment communication channel between
restricted hosts by taking advantage of allowed network traffic that passes through security software, effectively bypassing
the segmentation policies of the target network.

\begin{figure}
\centering
\includegraphics[width=0.8\columnwidth]{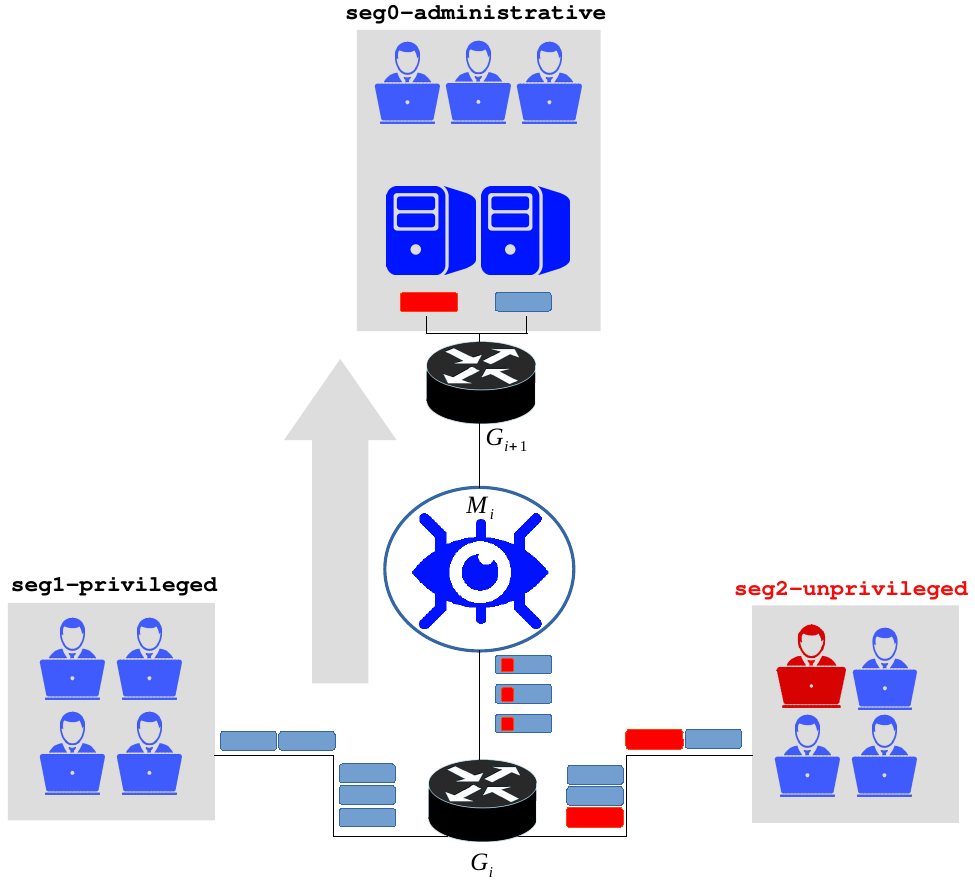}
\vspace{-0.2in}
\caption{Using CYPRESS we provide access to the administrative segment $seg0$ for the unprivileged segment 
$seg2$ }
\label{fig:segs}
\vspace{-0.15in}
\end{figure}

In this experiment, as shown in Figure~\ref{fig:segs}, we have three segments in the test network. The hosts in each segment 
are connected to a cGateway that connects the segment to monitor $M_i$ which centrally enforces access control and segmentation policies between all
the existing segments. Segment $seg0$ has two administrative
servers which are only accessible either locally by the users physically connected to the servers inside the segment or by users from
segment $seg1$. Any other packet sent to segment $seg0$ will be dropped by $M_i$. This includes packets from $seg2$ which is 
an unprivileged segment. Segments $seg1$ and $seg2$ are connected to cGateway $G_i$ and the administrative segment $seg0$ is connected
to cGateway $g_{i+1}$. Using a similar approach to the previous examples, we can hide the entire network traffic from a specific host in the
unprivileged segment inside packets communicated between $seg0$ and $seg1$. The hidden traffic from $seg2$ is extracted by $G_{i+1}$ 
and delivered to the correct host in $seg0$ after
the carrier packets pass through $M_i$ ; effectively creating a normal connection from
a restricted segment to the administrative segment.

\section{Detection and Mitigation}
\label{sect:detection_test}

Here, we evaluate how our data transmission approach resists detection and mitigation.
Detecting the presence of a covert channel in packets modified by CYPRESS depends on the ability to
detect the individual techniques used to create a covert channel in each separate protocol, not on the specific
networking hardware that routes the packets modified by CYPRESS. CYPRESS exploits the protocol, not any specific
software or hardware. Since the modified packets structurally conform
with TCP/IP protocol definitions (i.e. packets are not corrupted), the networking device treats them as regular 
packets and hence all networking hardware are exploitable\footnote{This includes the networking hardware on which
CYPRESS is running and the hardware that routes the modified packets both before and after the extraction stage.}.

In this section, we test standard well-known IDS tools not a specific
mitigation approach to detect a specific covert channel. 
Covert channel detection and prevention has not been exercised effectively in real-world network protection arsenal, generally because
the proposed techniques are complicated and expensive. The other reason that people
tend to overlook protecting networks against covert channels rests on two assumptions: 
\textit{covert channels are very rare} and
\textit{covert communication has a very limited capacity}. While there is no study to investigate to what extent covert channels
are used in real-world scenarios, we showed in this paper that the second assumption is not true. Also, CYPRESS is a framework
and is not limited to specific covert channels.
Note that a working technique (including any DPI
technique proposed in academia or industry) to detect a known
covert channel can also be effective in recognizing the channel if it is used by CYPRESS. For example, if a working
solution to identify a covert channel in TCP sequence numbers, might also work
to identify such a channel if CYPRESS uses TCP sequence numbers of some carrier packets for data transmission.
 
In the following we test a popular network anomaly detection solution, Zeek~\cite{zeek}, 
and a network analysis software, NetworkMiner~\cite{netwminer},
in different setups of gateways
and we report the result of packet analysis by these solutions. We send multiple payloads 
between gateways and run NetworkMiner
against the packets to check whether it can recover the secret payload. 
We also use known malicious files---like malware executables with known signatures--- as the secret payload and record all the packets
forwarded by CYPRESS after fusion. We then submit these packets to VirusTotal
to see whether it can detect any malicious file transferred by the carrier packets. VirusTotal is a world-renowned
\footnote{As of April 2023, VirusTotal receives more than two million submissions per day~\cite{vtstats}.}
online service that analyzes the submitted data using tens of security software. VirusTotal supports~\cite{vtpcap} scanning, 
recovery of transmitted files and in-depth security analysis of network data sent in
the form of PCAP files by investigating the packets using popular network intrusion detection systems like Snort~\cite{snort} and
Suricata~\cite{suricata}.

We run Zeek on a monitor placed between two cGateways. Zeek studies the structure and content of each packet that it captures and in case of an anomaly, it generates 
a log entry in one of its log files depending on the type and reliability of the potential anomaly. Zeek does
this by executing a large number of event-driven analysis scripts that are shipped with the tool. Each script is written
to find a specific incident that is potentially associated with an abnormal or dangerous case like various types of known network attacks. 

We start a transmission session on a secret node in which the secret user
downloads an arbitrary file from the Internet. We repeat the test multiple times changing the visible bandwidth and the number of visible users 
to evaluate their effect on the ability of Zeek to detect anomalies. Since CYPRESS distributes segments on different fields of carrier packets across
different connections, there is no general rule or configuration for any IDS that can accurately identify a packet
modified by CYPRESS. That said, increasing the number of users alone, makes detection more challenging.

\begin{figure}
\centering
\includegraphics[width=0.8\columnwidth]{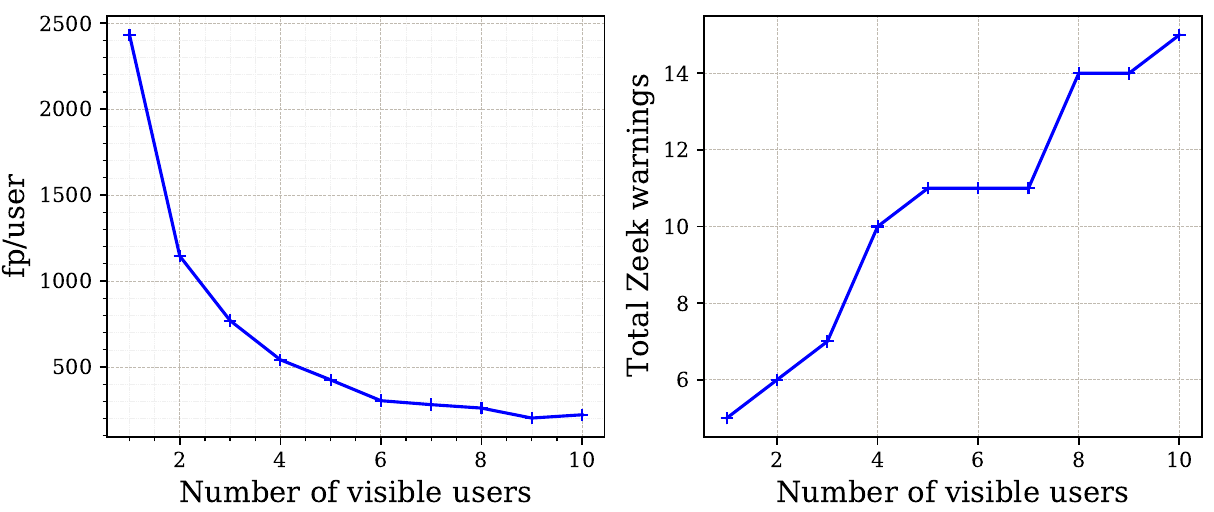}
\caption{Number of modified packets per user (left) and number of Zeek warnings (right).}
\label{fig:pckrep1}
\end{figure}

\begin{figure}
\centering
\includegraphics[width=0.9\columnwidth]{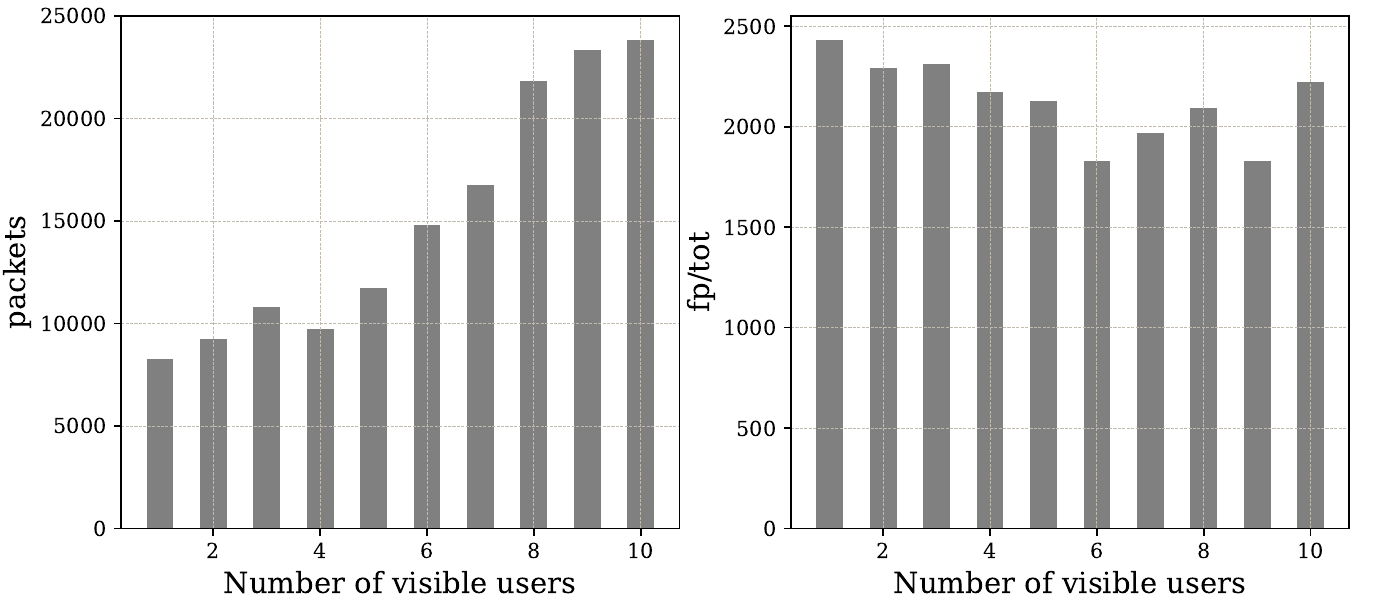}
\caption{Total number of packets passed through the gateways (left) and total number of modified packets (right).}
\label{fig:pckrep2}
\end{figure}

In Figure~\ref{fig:pckrep1} we have shown the number of modified packets to transfer the test file 
in a download session. The number of visible users in the sessions varied from 1 to 10. In each test, we studied the log
files of Zeek. The total number of warnings that Zeek generated is shown in Figure~\ref{fig:pckrep1}. 
In Figure~\ref{fig:pckrep2} we have reported the total number of modified packets along with the total number of packets that
passed through the gateways (\textit{fp} stands for fusion-packets).

While it is possible to detect a covert channel by investigating the reported anomalies, we argue that in most situations, 
it is not a strong determining factor to prove the presence of a covert channel for three reasons. First, 
the reported
anomalies can also contain false positives in the case of benign connections. There are various cases in which
Zeek might report a potential problem/anomaly while there hasn't been any notable or real issue
in the communication. Second, even in the case of a network problem, there exist many reasons that could have caused it. Some of the
possible reasons 
are not security-related. For example, the source of the issue can stem from common network problems like disconnection, 
packet loss, or 
retransmissions. Even for security problems, a network anomaly does not necessarily determine that a covert channel has been
in place, given that the description of each detected anomaly (in Zeek and many other security solutions) is quite generic.
Third, the rate of the anomalies reported (even without excluding potential false positives) is insignificant compared to the total 
number of carrier packets that leave a cGateway. In none of our tests could Zeek reliably find a potentially abnormal/malicious stream
of packets. Note that even if the detection
can be successful, CYPRESS as a framework can be adapted to use new techniques and channels that are not supported by these tools.

We repeated the transmission sessions with known malware executables and 
recorded the packets in all sessions, each containing at least an un-encrypted malicious executable with a known signature and ran NetworkMiner to analyze
the stored PCAP files. NetworkMiner is a network forensic analysis tool that can accurately parse network packets to
reassemble and recover transmitted files. We ran NetworkMiner against all the recorded PCAP files for each 
test session separately. NetworkMiner failed to recover the transmitted file in every test. Similarly, uploading the
PCAP samples to VirusTotal resulted in zero malicious file detections for every submission. 
Note that the sessions were executed with \textit{disabled} encryption and still, all the tests in VirusTotal failed to 
recover any of the files that were hidden in the packets and consequently failed to detect a single malicious PCAP file. 
This shows that while we can reliably transmit secret data between nodes, it is extremely challenging for security solutions 
to recover the transmitted secret payload from the carrier packets of legitimate users.

\section{Future Work}
We used a small set of covert techniques in CYPRESS. As mentioned earlier,
there are many more possible covert channels that can be compiled into the framework.
We will work on incorporating more covert channels
in CYPRESS. Our current prototype supports
multiple storage-based techniques;
we may add support for timing channels in the future.

Covert channels can both help and threaten anonymity networks.
It is possible to use covert channels to deanonymize users behind an anonymity network.
While the application data layer is protected from tampering using encryption in anonymity networks,
lower layers, particularly the transport and IP layers, can convey covert data. Some of the 
network fields in these layers are preserved between routers and are recoverable 
later on along the route. Coordinating the covert data at different routers can potentially 
compromise the user's anonymity. This is a separate area of research that
we consider as a future project.
Finally, we will study more effective solutions to identify covert communication in
 different layers as the current anomaly detection methods fall short in 
accurate detection of carrier packets used by CYPRESS.

\section{Related Work}
There are many papers that discussed covert channels. However, 
they have only focused on using a specific covert transmission technique or exploring 
and comparing different techniques as a survey. ~\cite{rowland} introduces 
using IP \textit{id} field and TCP sequence numbers. 
~\cite{cabuk} uses a timing channel where delays are placed between successive 
transmissions and then read by the other side. ~\cite{cauich} uses \textit{fragmentation offset} and 
\textit{identification} fields in the IP header. ~\cite{Mazurczyk} 
uses Path MTU Discovery (PMTUD) and  Packetization Layer Path MTU Discovery (PLPMTUD) to 
implement a covert channel by interpreting the number of fragments and fragmentation 
rate at the receiver side. ~\cite{qu} used TTL field of the IP header. 
~\cite{gianvecchio} and ~\cite{Sellke} use statistical 
models in IPDs to transmit covert data at a low bit rate. 
~\cite{Trabelsi,Trabelsi2,Trabelsi3} use trace routing and IPv4 RR 
(Record Routing) option. ~\cite{graf} uses IPv6 options header. 
Several sources have discussed carrying IP packets using ICMP ~\cite{icmp1,icmp2,icmp3,icmp4,icmp5} 
and ~\cite{igmp} introduces IGMP-based covert channels. Other protocols like DHCP ~\cite{rios} and ARP~\cite{Liping} 
have also been studied for this purpose.

These are all effective stealth communication techniques but with extremely limited use. 
Despite the high number of research works on 
network-based covert communication methods, 
there has not been any prior work to implement a comprehensive framework to incorporate 
and synchronize different types of covert channels to provide a reliable connection between 
network nodes by manipulating legitimate traffic of visible users. To the best of our knowledge,
CYPRESS is the first framework with this capability.

\section{Conclusion}
Real-world capabilities of covert communication have been under-explored. 
Some prior works generate new packets to transmit covert data, and many of the proposed techniques have extremely 
limited practical use. We showed that multiple 
connections can be covertly stacked on top of one single stream of 
packets to provide a communication strategy that is hard to detect and stop. 
Security solutions are not designed to monitor and understand unconventional ways
of data transmission which makes it possible to move large amounts of disallowed data around
advanced monitoring tools.
The security community needs to expand its research 
and arsenal to defeat covert communication which can be used to
hide malicious activities. Covert channels can also provide 
anonymous Internet access to legitimate users that 
need to hide their network activities from an adversary. This can complement the existing
anonymity services like the Tor project.

\bibliographystyle{IEEEtranS}
\bibliography{IEEEabrv,cypress}

\appendix
\section{Assigning The Costs}

Cost assignment for a new covert channel is performed based on the experimental results of using 
that covert channel alone for the transmission of a group of sample packets, over multiple constant
transmission sessions, and then calculating the average value of the delays and comparing it
with the minimum and maximum thresholds.

The thresholds are calculated for each transmission session independently. The minimum threshold, \textit{MIN\_TH}, is 
the time that it takes for the transmission session to complete for a visible
user when no covert transmission is in effect. This is the normal communication time when CYPRESS is 
not used. In other words, \textit{MIN\_TH} shows the transmission time when there is no cost
involved for covert transmission. Maximum threshold, \textit{MAX\_TH}, is the time that it takes to complete the transmission session
when CYPRESS is configured to sporadically use the covert channels with augmented correction with a 
constant probability that does not slow down the secret communication to less than 100KB of bandwidth when
one visible user is used. Similarly, when we have more than one visible user, this minimum speed roughly aligns with 
half of the corresponding number 
reported in Figure~\ref{fig:spdbw}. Note that all these configurations are \textit{experimnetal} 
and there is not a certain specific value for either of these thresholds. 

\begin{figure}[h]
\centering
\includegraphics[width=1\columnwidth]{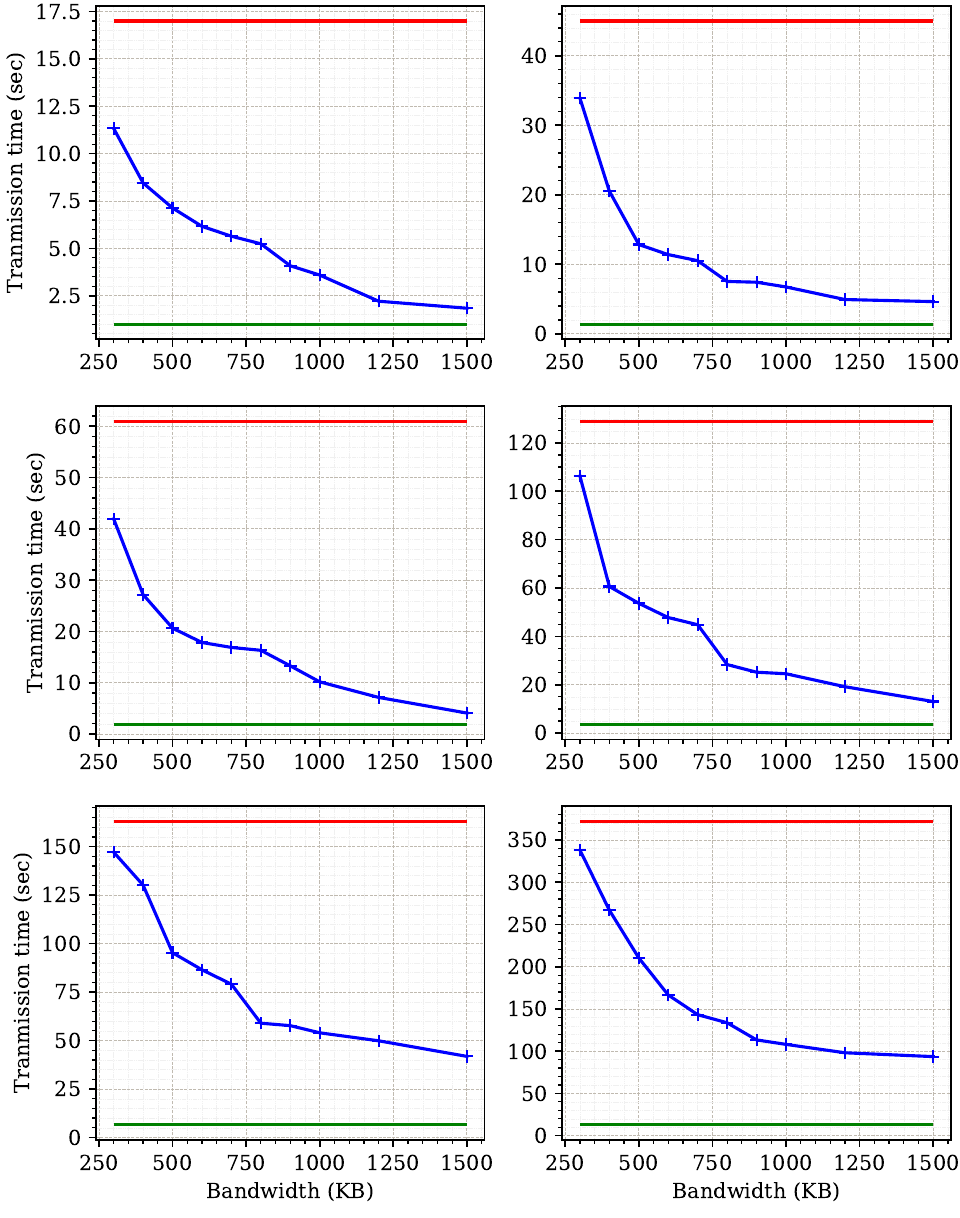}
\caption{Sample cost assignment in 6 transmission sessions and 60 rounds. The horizontal green
and red lines in each plot is MIN\_TH and MAX\_TH respectively.}
\label{fig:cost}
\end{figure}

We execute at minimum six different transmission sessions, each of which configured to transfer certain
amounts of secret data, while gradually increasing the transmission size with the session number
progressing. Each session is run with three visible users and ten different bandwidth values 
for each visible user. One secret user is connected to the test network. The cost of the channel is calculated 
as the average cost of all the transmission sessions as written below.

$$f_C=\sum_{i=1}^{n}\frac{t_i-MIN\_TH}{n(MAX\_TH-MIN\_TH)}$$

where $t_i$ is the time that it takes to complete the transmission in test $i$, and $n$ is the number of the tests. $t_i$ is
always less than $MAX\_TH$.

In Figure~\ref{fig:cost} we have shown the result of evaluating one specific covert channel (TCP options) in the six transmission
sessions that have been used for the channel. In this test $n=60$ as we have total of 60 tests (six sessions and ten bandwidth
levels). $f_C$ is calculated as $0.34$ in this test which is the collective cost of synchronization complexity and network
slow-down. For all candidates that are reserved to hold synchronization headers, the cost must be less than $0.05$ to sustain the
communication stability. Note that we used constant thresholds for different bandwidth in this example. In the case of variable
$MIN\_TH$ and $MAX\_TH$ for each bandwidth, we will have:

$$f_C=\sum_{i=1}^{n}\frac{t_i-MIN\_TH_i}{n(MAX\_TH_i-MIN\_TH_i)}$$

\label{ap:costs}

\section{Algorithms}
\label{ap:alg}

In Algorithm \ref{alg:algfusion} and Algorithm \ref{alg:algextract}, we have written a high-level pseudocode for
the fusion and extraction stages used in CYPRESS. 

\begin{algorithm}[!h]

\caption{CYPRESS-FUSION}
\label{alg:algfusion}
\SetAlgoNoLine
\SetKwInOut{Input}{Input}
\Input{Current state of PMS as $hfn$}
\While {true}{
 $pck$ = recv\_next\_packet()\;
 \If {(from\_a\_secret\_client($pck$))}{
 	queue\_secret\_packet($pck$)\;
 	continue\;
 }
 \If { ($pck\_state$=verify\_fusion\_path($pck$)) }{
 	\If {(!transmission\_queue\_ready())}{
 		$secret\_pck$=pop\_next\_secret\_pck($sec\_pck\_queue$)\;
 		update\_transmission\_queue($sec\_pck\_queue$)\;
 	}
 	$tq\_reference\_pck$=load\_transmission\_queue();
 	\If {($secret\_pck$)}{
	 	$cnd\_offsets$ = PMS\_evaluate\_pck($pck$,$pck\_state$,$hfn$)\;
	 	
	 	\ForEach{ $hnd$ in $cnd\_offsets$}{
	 		$nf$=next\_frag($tq\_reference\_pck$,$hnd$)\;
	 		
	 		\If {($encrypt\_frags$)}{
	 			$nf$=encrypt($nf$)\;
	 		}
	 		
	 		\If {(is\_initial\_in\_session($nf$))}{
	 			$sync\_hdr$=generate\_synh($nf$,$hnd$)\;
	 			$nf$=prepend\_synh($nf$,$hnd$)\;
	 		}
	 		
	 		fusion\_per\_hnd($hnd$,$pck$,$nf$)\;
	 		
	 	}
 
 	}\Else{
 		exclude\_pck(pck);
 	}

 }
 
 $pck$=route\_l3($pck$)\;
 $pck$= update\_phys($pck$)\;
 
 retransmit\_phys($pck$)\;

 }
 
\end{algorithm}

\begin{algorithm}[!h]

\caption{CYPRESS-EXTRACTION}
\label{alg:algextract}
\SetAlgoNoLine
\SetKwInOut{Input}{Input}
\Input{Current state of PMS as $hfn$}
\While {true}{
 $pck$ = recv\_next\_packet()\;
 
 \If { ($pck\_state$=verify\_extraction\_path($pck$)) }{
 	$sec\_pck\_ref$=load\_recv\_buffer()\;
 	$cnd\_offsets$ = PMS\_evaluate\_pck($pck$,$pck\_state$,$hfn$)\;
 	\ForEach{ $hnd$ in $cnd\_offsets$}{
	 		$nf$=extract\_per\_hnd($hnd$,$pck$)\;
	 		$sync\_hdr$=evaluate\_sync($nf$)\;
	 		\If {($syn\_hdr$)}{
	 			synchronize\_recv\_state($nf$,
	 				$sync\_hdr$,$sec\_pck\_ref$)\;
	 		
	 		}
	 		
	 		\If {($decrypt\_frags$)}{
	 			$nf$=encrypt($nf$)\;
	 		}
	 		
	 		load\_frag\_in\_buf($nf$,$sec\_pck\_ref$)\;
	 		
	 		\If{(local\_buffer\_completed($sec\_pck\_ref$))}{
	 			$sec\_pck$=recover\_packet\_whole($sec\_pck\_ref$)\;
	 			$sec\_pck$= update\_phys($sec\_pck$)\;
 
 				retransmit\_phys($sec\_pck$)\;
	 		}
	 		
	 }

 $pck$=route\_l3($pck$)\;
 $pck$= update\_phys($pck$)\;
 
 retransmit\_phys($pck$)\;

 }
}

\end{algorithm}


In summary, this is what happens in the fusion stage
to write secret segments into the packets sent from $v_i$ (again, gateways are symmetric and what we describe for one 
route holds for the opposite route as well):

\begin {compactitem}
\item 
The transmission direction and physical addresses are checked to verify that the packet belongs to 
the correct path of fusion.

\item CYPRESS looks for potential storage candidates in each of the packets passing through 
\footnote{Destructive carrier fields like header checksum are fixed in a next cGateway
before forwarding the packets to the next regular gateway.}.

\item Based on the handler function defined for each potential field and the state of the connection, some of the fields of the current
packets are selected to carry secret segments. Some handlers are only limited to specific connection states. For example, we
do not use TCP options for SYN packets. By default, only one of the possible handlers is used per packet.

\item Secret data is fragmented based on the usable size of the chosen candidates. A small synchronization header is prefixed
 to the first segment to instruct the next cGateway on how to recover the secret data by using the correct handler and fix the destructive fields to
 prevent corrupting the packets when they arrive at the destination node, which is not running CYPRESS.
 
\item If encryption is enabled, secret segments are encrypted using AES-256 and a symmetric key that is only known to
the cGateways.

\item The carrier packet is passed to the chosen handler functions to go through necessary modifications.
\item The final content of the modified values is prepared for transmission to the next physical node.
This might include translating the local IP addresses (NAT procedure) if the response packet should go through
a protected path. Finally, the packet is sent to the NIC.
\end {compactitem}

In the extraction stage, which happens in the receiving cGateway, we have the following steps:

\begin {compactitem}
\item The transmission direction and the physical addresses are checked to verify the packet belongs to 
the correct path of extraction.
\item The packet structure is studied and the presence of a potential synchronization
header is checked using header signatures\footnote{The signatures or synchronization header's magic bits
are defined for CYPRESS at compile time.}, and if found, the local storage buffer and low-level extraction counters that specify
the progress of recovered segments, are updated
accordingly. Based on the chosen handler functions and synchronization instructions, secret fragments are
recovered from the packet. 
\item If encryption is enabled, the segment is decrypted after we have extracted enough cipher blocks for
stream deciphering.
\item The recovered segment is stored in the local buffer. If this is the last segment, the secret packet is 
formed and prepared for transmission.
\item Potential destructive fields in the carrier packets are fixed and overwritten and the packet is prepared for
transmission; ending up in the NIC to leave the cGateway and is forwarded to the next correct gateway.

\end {compactitem}

\section{Example Handler Function}
\label{ap:hnd_example}

We describe the ICMP echo request handler in this example. This is relatively a short and simple 
handler since no network options are involved. For these packets, the protocol type of the IP header
is one, denoting ICMP protocol at the bottom layer. PMS is segmented based on the protocol type 
and hence the correct handler from the appropriate segment is loaded. The handler signature is defined in Listing~\ref{lst:example}.

\textit{pck} points to the first byte of the link layer.
ICMP\_REQ is defined as eight. ETHER\_SIZE and IPH\_SIZE are the sizes of the link layer and the IP 
header, respectively (for the cases that have IP options, the size is adjusted accordingly). The 
moderation code receives the packet and the secret segment (fragment in Algorithm~\ref{alg:algextract}) chosen by the core fusion code. 
If the secret segment
is not ready for transmission, the packet is \textit{excluded} from the fusion and an exclusion signature is 
added to the packet. Otherwise, the storage candidate starts from the end of the IP header plus 8 bytes which
will be the first byte of ICMP optional payload after the ICMP header. Standard Linux ping program sends 56-byte
payloads with the initial 8 bytes set to the current timestamp. Based on whether the timestamp should be preserved
or not, the handler chooses the start offset of the write operation and then proceeds to write as many bytes as possible
in the ICMP payload. The pseudo-code of this operation is written in Algorithm~\ref{alg:icmp}.

\begin{lstlisting}[caption=Example signature for ICMP echo request,label=lst:example]
(struct udph*)(pck + ETHER_SIZE + IPH_SIZE)->type==ICMP_REQ
\end{lstlisting}

\begin{algorithm}[h!]

\caption{ICMP-moderator}
\label{alg:icmp}
\SetAlgoNoLine
\SetKwInOut{Input}{Input}
\Input{$pck$, $nf$}

\If {(!fragment\_ready($nf$))}{
	exclude\_pck($pck$);
}
$fusion\_offset$=$pck$ + ETHER\_SIZE + iph\_size($pck$) + ICMPH\_SIZE\;
\If {(ICMP\_preserve\_ts)}{
	$fusion\_offset$+=ICMP\_TS\_SIZE;
}
$nw$=frag\_raw\_update(pck,$nf$,$fusion\_offset$)\;
fix\_checksum($pck$)\;
return $nw$\;
\end{algorithm}

\section{Impersonating Visible Users}
\label{ap:hijack}

In the previous experiments that we discussed in the paper, we carried the network traffic of
secret users on the visible users' network traffic. 
The basic idea was to hide a secret user
from the monitor by preventing the monitor from \textit{seeing} the secret user's packets. On the other hand,
we can allow the monitor to see the secret packets but think of them as legitimate packets belonging 
to a visible user. In other words, 
we can reverse the previous policy to have the secret
user look like one of the legitimate visible users while having the target visible user still use his 
correct IP address.

\begin{figure}[ht]
\centering
\includegraphics[width=0.6\columnwidth]{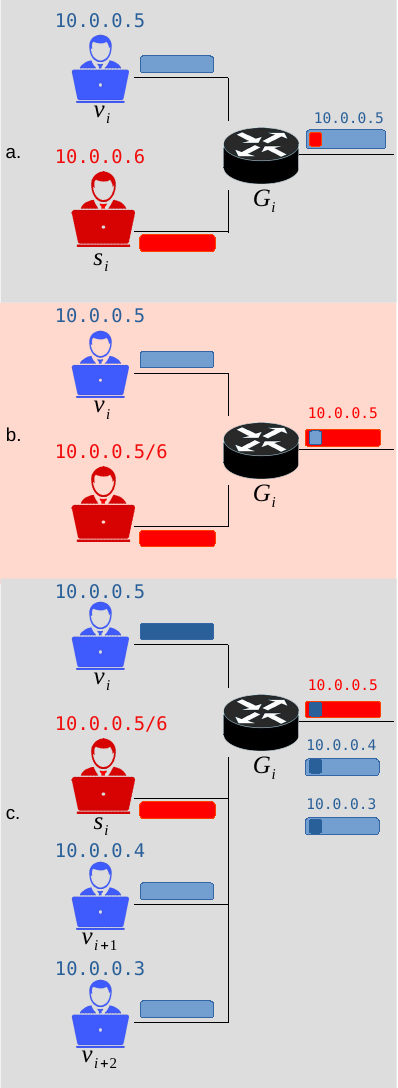}
\caption{\textbf{a.} Secret packets from $s_i$ are carried through packets of visible user $v_i$. \textbf{b.}
Secret packets are passed directly after forging $v_i$'s IP address, $v_i$'s packets are carried through 
secret packets. \textbf{c.} $v_i$'s packets are distributed on all users' packets.}
\label{fig:switch}
\end{figure}

In Figure~\ref{fig:switch} the visible user $v_i$ is given an IP address of 10.0.0.5
and the secret user $s_i$ is assigned 10.0.0.6 which is only known to the gateway $G_i$ and optionally
a DHCP server that assigns IP addresses to the clients. Although in our threat model, the secret users should be only
hidden from the monitor, as we mentioned before, since secret users are behind 
a cGateway, their IP addresses need not to be exposed to any node other than the cGateway that is directly connected
to the secret users. CYPRESS can assign any arbitrary address to a secret user irrespective of the entire network environment
and other visible users. This means that we can even use the IP address of a connected visible user like 
$v_i$ and assign it to $s_i$. Similarly, we can change the IP address of outgoing packets
from $s_i$ to any arbitrary address and later translate the destination IP address of the 
response packets before delivering them to $s_i$; the arbitrary IP assignment can be done
both before and after the packet leaves the secret user's machine. Note that CYPRESS knows the physical (MAC) addresses of the connected hosts, and that 
suffices to differentiate
a visible and a secret user for an outgoing packet. For the incoming packets that should be delivered to
the users behind the cGateway, we use an unconventional form of Network Address Translation by mapping
the source port of the outgoing packet to the \textit{physical addresses} of the users that are directly connected
to the cGateway instead of mapping the source ports to the IP addresses. This way, we can reuse an IP address for
multiple hosts at the gateway level.

In this example, we assume that the monitor only allows
traffic from $v_i$ with the IP address 10.0.0.5 to pass through. In the previous experiments, we mounted $s_i$'s traffic on top of 
$v_i$'s packets (Figure~\ref{fig:switch}-a). This is the normal scenario in which the packets that
are passed through by the monitor belong to $v_i$. However, we can carry out the opposite procedure
and mount the network traffic of $v_i$ on top of the network traffic of $s_i$ with one important difference:
we assign the IP address of $v_i$ (10.0.0.5) to $s_i$ but let $v_i$ keep using its original IP address
(Figure~\ref{fig:switch}-b). From this moment on, the secret user $s_i$ is directly connected to the network and can communicate
with outside like a visible user, and at the same time, $v_i$ fails to notice that its role has been changed
to a secret user. The monitor still sees the network activity of \textit{one} user -which is perceived to be $v_i$- going through
the route, while the packets that the monitor sees belong to $s_i$.

Whatever $s_i$ does will be seen and logged as the network activity of $v_i$. The communication between
$v_i$ and outside is provided through $s_i$ \textit{and} potentially any other visible user behind $G_i$.
This means that the connection speed goes up for $s_i$ and goes down for $v_i$. The new bandwidth that
is provided for $v_i$, similar to what was described before for secret users, will depend on the number of visible
users and the existing bandwidth. This means that when there are multiple visible users behind a cGateway,
at any moment in time, CYPRESS can have a secret user switch its role with a target visible user, effectively
hijack his IP address and impersonate the target user(Figure~\ref{fig:switch}-c). However, this scenario has 
an important drawback. The security software can monitor the secret packets and identify malicious activity.
The transmitted files can be recovered and scanned, and therefore, the results discussed in section~\ref{sect:detection_test}
will not be possible. That said, the goal of the explained strategy in this example is to impersonate random users, not to hide 
the network activity or create secret Internet access. This is an example of how we can surreptitiously 
switch the roles of users within a local network environment only by using covert channels that exist in almost
all network packets.

\end{document}